\documentclass[twocolumn,groupedaddress,amsmath,amssymb,prl,aps,longbibliography]{revtex4-1}
\usepackage{graphicx}
\usepackage{dcolumn}
\usepackage{bm}
\usepackage{booktabs}

\usepackage{tikz}
\definecolor{mydarkgreen}{rgb}{0.0, 0.5, 0.0} 
\usepackage{rotating}
\usepackage{gensymb}
\usepackage{xcolor}
\newcommand{\bfo}{BiFeO$_3$}
\newcommand{\lno}{LaNiO$_3$}

\begin{document}

\title{Engineering ferroelectric nanocapacitors through interfacial phonon bias}

\author{I\~nigo Robredo-Magro,$^{1}$ Urmimala Dey,$^{1}$ Suehyun
  Park,$^{2}$ Karam Cho,$^{2}$ Raseong Kim,$^{2,3}$ and Jorge
  \'I\~niguez-Gonz\'alez$^{1,4}$}

\affiliation{ \mbox{$^{1}$Smart Materials Unit, Luxembourg Institute
    of Science and Technology (LIST),} {Avenue des Hauts-Fourneaux 5,
    L-4362 Esch/Alzette, Luxembourg}\\ \mbox{$^{2}$ Technology Research Intel Foundry, Hillsboro, Oregon 97124,
    USA}\\ \mbox{$^{3}$Computational and Modeling Technology, Intel Foundry, Hillsboro, Oregon 97124, USA}\\ \mbox{$^{4}$Department of Physics and Materials Science,
    University of Luxembourg, Rue du Brill 41, L-4422 Belvaux,
    Luxembourg }}

\begin{abstract}
As ferroelectric devices approach the nanometer scale, their behavior
becomes increasingly sensitive to boundary conditions. In perovskite
oxide nanocapacitors, epitaxial strain and electrostatic screening are
well-established control parameters. Then, while expected to
contribute to the epitaxial stabilization of ferroelectric ultrathin
films, the lattice-dynamical state of the electrodes has not been
isolated as an independent tuning knob. Here we call this poorly
explored coupling \emph{interfacial phonon bias}, whereby the
electrodes' phonon energy landscape reshapes that of the adjacent
ferroelectric across their shared interface. Using
first-principles-accurate machine-learned interatomic potentials and
molecular-dynamics simulations, we examine this mechanism in symmetric
BiFeO$_{3}$ nanocapacitors in contact with a metallic perovskite
oxide. By independently controlling the orientation and amplitude of
the electrodes' octahedral-tilt landscape, while keeping conventional
elastic and electric boundary conditions fixed, we show that the
electrodes' phonon energy landscape can modify the stability of
competing BiFeO$_{3}$ phases. In particular, this allows us to shift
the ferroelectric transition temperature and even induce
antiferroelectric-like switching at ambient conditions. We thus
showcase interfacial phonon bias as a complementary and powerful
strategy to modify and optimize the properties of ultrathin epitaxial
films.
\end{abstract}
\maketitle

Boundary conditions govern the behavior of ultrathin films of
functional materials, ferroelectric oxides being a notable example. In
thin films and nanocapacitors, the best-established control parameters
for epitaxial stabilization are epitaxial strain, which reshapes the
structural free-energy landscape through lattice mismatch, and
electrostatic boundary conditions, which determine the screening of
the spontaneous polarization and the associated depolarizing
fields~\cite{pertsev98,schlom07,dawber05,stengel06,lichtensteiger14}. These
two effects have been studied extensively and are now central to the
design of ferroelectric
heterostructures~\cite{dawber05,schlom07}. However, epitaxial
stabilization is not governed by elastic and electric boundary
conditions alone: the structural connectivity across a coherent
interface can also modify the relative stability of competing
phases. Yet, the internal structure and lattice-dynamical features of
the surrounding oxide environment have remained largely unexplored as
an independent control knob.

The ability to independently control structural boundary conditions is
especially valuable in ferroelectrics with competing polymorphs, where
relatively small changes in the free-energy landscape can select
polar, antipolar, or nonpolar
states~\cite{rabe13,akansha18,dieguez11}. Such phase selection can
determine, for example, whether a nanocapacitor exhibits conventional
ferroelectric switching~\cite{liu16a,Shi2022}, a field-induced
transition from an antipolar to a polar
state~\cite{kittel51,graf14,lu20}, or a more complex multistep
polarization reversal~\cite{stengel15,fedorova22}. It can therefore
modify key functional properties, including the switching barrier and
coercive field~\cite{liu16a,Shi2022}, dielectric and magnetoelectric
responses~\cite{fiebig16,wojdel09}, and energy-storage
behavior~\cite{xu17,aramberri22}. A structural control parameter that
operates through the electrodes' internal structure and phonon energy
landscape, rather than through elastic mismatch or electrostatic
screening, thus provides an additional route to engineer phase
stability and switching behavior in nanoscale ferroelectric devices.

\begin{figure}[t]
    \centering
    \includegraphics[width=1\linewidth]{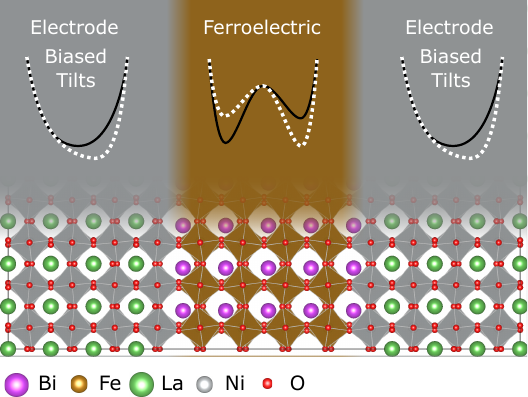}
    \caption{Concept of interfacial phonon bias and symmetric
      nanocapacitor architecture. Solid black and white dashed lines represent the intrinsic phonon energy landscape and the biased one, respectively. Electrode: the octahedral-tilt
      energy landscape is modified to favor a selected structural
      state. Interface: the shared interface (e.g., oxygen-octahedral
      network in oxide perovskites) provides the pathway for
      structural coupling across layers. Ferroelectric: the modified
      electrodes' landscape reshapes the free energy and phase
      competition of the nanocapacitor.}
    \label{fig:fig1}
\end{figure}

This mechanism can be especially important in heterostructures of
structurally compatible materials, where structural distortions can
couple across the
interface~\cite{PhysRevLett.105.087204,Lu2015,PhysRevB.82.113402,PhysRevLett.105.227203}.
In perovskite oxides, for example, the corner-sharing
oxygen-octahedral network provides a direct link between the
structures of adjacent
layers~\cite{zubko12,bousquet08,Rondinelli2012}.  The importance of
this connectivity is illustrated by the pronounced growth-orientation
dependence of the metal-to-insulator transition temperature in
NdNiO$_{3}$ films on
NdGaO$_{3}$~\cite{Catalano_2018,10.1063/1.4919803}, and by the
suppression of oxygen-octahedral breathing distortions reported in
BaBiO$_{3}$-based
heterostructures~\cite{PhysRevB.78.132507,10.1063/1.5010825}. In these
systems, however, interfacial structural coupling is generally
intertwined with epitaxial strain, confinement, interface chemistry,
and electrostatic effects, making its individual contribution
difficult to isolate. Here, we use predictive atomistic simulations to
isolate and explore this mechanism, which we call \emph{interfacial
phonon bias} (see Fig.~\ref{fig:fig1} for a schematic
representation). Our results show it provides us with a powerful knob
to engineer the behavior of ferroelectric nanocapacitors.

This effect may be particularly relevant for materials like \bfo,
where polarization is strongly coupled to oxygen-octahedral
rotations~\cite{dieguez11,gu18,fedorova22}, so that changes in the
tilts' free-energy landscape are expected to substantially affect the
behavior of the ferroelectric polarization. Consistently, recent
experimental and theoretical work has shown that switching in
\bfo\ cannot be understood solely in terms of polarization and strain:
the evolution of octahedral tilts during reversal is essential to the
switching barrier and kinetics~\cite{heron14,Shi2022,fedorova24}. The
competition between the polar $R3c$ and antipolar phases in \bfo\ has
also attracted considerable interest as a possible route to achieve
antiferroelectric-like behavior and energy-storage
functionality~\cite{dieguez11,cazorla13,graf14,doi:10.1126/sciadv.abg5860}. Controlling
this $R3c$--$Pnma$ phase competition through the interfacial phonon
bias would therefore offer an alternative to the strain-,
electrostatic- and composition-based approaches commonly used to
stabilize competing nonpolar or antipolar
phases~\cite{infante10,kan10,gonzalezvazquez12}.

In this work, we isolate the interfacial-phonon effect by design. We
choose \lno\ perovskite electrodes whose interfaces with \bfo\ do not
involve any nominal-charge discontinuity that could obscure the
structural physics of interest. We work with symmetric
\bfo\ nanocapacitors, considering the NiO${}_2$/BiO composition for
both interfaces and in all calculations, so that the electric boundary
conditions are kept constant. The in-plane lattice parameter is fixed
to that of a canonical substrate, i.e. SrTiO$_{3}$ oriented along the
$[001]_{\rm pc}$ pseudocubic direction, to impose the epitaxial
constraint. (See Fig.~\ref{fig:fig1} and Supplemental for full
structural details.) We then tune the electrodes' structural state by
controlling the orientation, amplitude, and stiffness of its
octahedral tilts.

We track the antiphase and in-phase octahedral-tilt amplitudes about
the pseudocubic axes by $R_{i}$ and $M_{i}$, respectively
($i=x,y,z$). We quantify these and relevant antipolar distortions
$A_{i}$ \cite{chen18a} ($i=x,y,z$ being the direction of the
anti-phase modulation) by projecting atomic displacements onto
symmetry-adapted modes of the cubic perovskite reference structure. We
also track an approximate polarization $P_{i}$ computed using nominal
charges. Full definitions and normalization conventions are provided
in the Supplemental.

To isolate the role of the electrodes' tilt energetics, we require a
way to vary its preferred tilt pattern and stiffness without changing
its chemistry or the epitaxial and electrostatic boundary
conditions. A so-called {\it harmonic bias} provides this control
while allowing the distortions of the electrodes and the ferroelectric
to relax together. Following similar approaches found in the
literature~\cite{PhysRevB.88.094101,MARAGLIANO2006168}, we add a
harmonic bias term to the electrodes' tilt degrees of freedom of the
form
\begin{equation}
E =k\sum_i \left(Q_i-Q^0_i\right)^2,
\end{equation}
where $Q_i\in\{R_x,R_y,R_z,M_x,M_y,M_z\}$, $Q^0_i$ are the target
amplitudes, and $k$ is the stiffness (see Supplemental for full
details on our implementation of the harmonic bias). To simulate these
nanocapacitors at realistic sizes (thousands of atoms) and over
timescales beyond the reach of direct first-principles methods (few to
tens of nanoseconds), we employ a machine-learned interatomic
potential based on the neuroevolution potential
formalism~\cite{fan21}, trained on first-principles data for
\bfo/\lno\ heterostructures and individual bulks (see Supplemental for
validation of the model).

We engineer our model electrodes in the following way. First, we
flatten the native \lno\ tilt energy landscape: we add a harmonic
compensating term that counteracts the intrinsic instability of the
\lno's antiphase tilts, bringing the electrodes close to the threshold
between untilted and tilted configurations. We estimate a critical
compensating stiffness of $k_{\rm comp}=4$ eV/(\AA rad), consistent in
magnitude with the quadratic coefficient obtained from Density Functional Theory (DFT) for the
Landau model description of bulk \lno\, $\kappa_r=-5.4$~eV/(\AA
rad)~\cite{chen18a}.

A second harmonic term then defines a preferred finite-amplitude tilt
pattern in our model electrodes. We select the $a^-b^+a^-$ tilt
configuration in Glazer's notation~\cite{glazer75}. Specifically, we
use the $a^-b^+a^-$ structure of bulk SrRuO$_{3}$ (SRO) as a relevant
reference, preserving the relative amplitudes of the antiphase and
in-phase rotations while varying their overall magnitude. The full
bias term thus reads
\begin{equation}
\label{eq:ebias}
E_{\text{bias}} = k_{\rm comp}\sum_i Q_i^2+k_{\rm tg}\sum_i
\left(Q_i-f_QQ_i^{SRO}\right)^2,
\end{equation}
where $f_Q$, $k_{\rm tg}$ are a scaling factor controlling the electrodes' target
tilt amplitudes and target stiffness, while $Q_i^{SRO}$ are the equilibrium tilt amplitudes
of bulk SRO (i.e., $R_x=R_z\approx 6.6^\circ$ and
$M_y\approx7.4^\circ$, all other tilt components being zero). This is
a particularly relevant octahedral tilt pattern because it corresponds
to the most common one found in perovskites~\cite{lufaso01,chen18a},
and in particular to the antipolar $Pnma$ phase of \bfo. It is
therefore very common among potential perovskite electrodes and well
suited to bias the competition between the ferroelectric $R3c$ and the
antipolar $Pnma$ phases.

\begin{figure}
    \centering
    \includegraphics[width=1\linewidth]{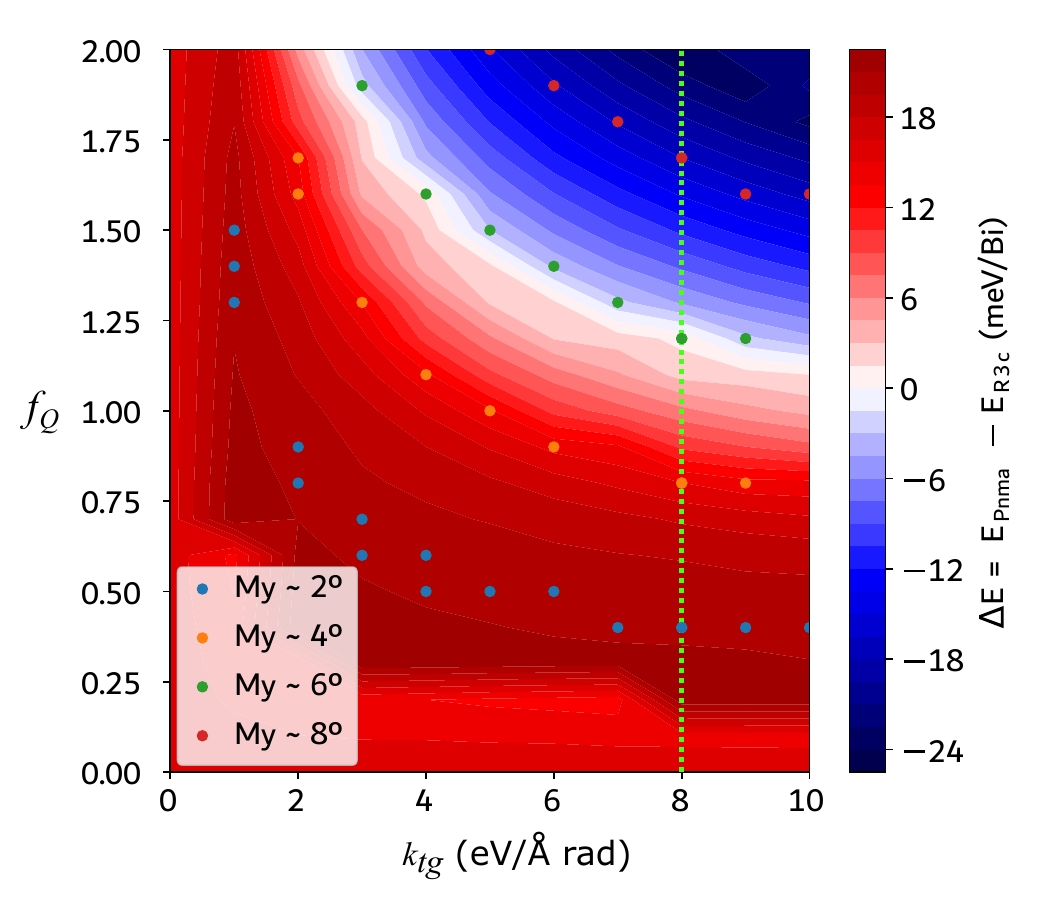}
    \caption{Energy difference between the $Pnma$ and $R3c$ polymorphs
      of a 3.2~nm \bfo\ nanocapacitor as a function of the stiffness
      ($k_{\rm tg}$) and target
      amplitude ($f_{Q}$) defining the harmonic bias applied to the
      electrodes. The scattered points mark regions where the
      resulting electrode tilt amplitude ($M_{y}$) is constant. The
      green dashed line marks the cases explored in
      Fig.~\ref{fig:fig3}.}
    \label{fig:delta_E}
\end{figure}

We start our analysis by computing the energy difference between the
$Pnma$ and $R3c$ polymorphs for the \bfo\ nanocapacitor, $\Delta
E=E_{Pnma}-E_{R3c}$, which is a proxy for the transition
temperature~\cite{cazorla13,jinnouchi19}, as a function of the scaling
factor $f_Q$ and stiffness $k_{\rm tg}$. Figure~\ref{fig:delta_E}
shows the resulting landscape, where we mark points at which the
electrodes exhibit the same in-phase tilt amplitude after structural
relaxation of the entire nanocapacitor (the antiphase tilts follow the
same trend). The target amplitudes $f_QQ_i^{SRO}$ define the preferred
electrode tilt pattern, while the stiffness $k_{\rm tg}$ sets the
energetic penalty for deviations from it, as given in
Eq.~\ref{eq:ebias}. The relaxed electrode amplitudes are therefore not
imposed directly, but emerge from the balance between this bias, the
intrinsic electrode energy landscape, and the interfacial coupling to
the \bfo\ layer. The lines of constant $\Delta E$ are approximately
parallel to those of constant $M_y$, indicating that the realized
average electrode tilt amplitude is the key structural variable
controlling the relative stability of \bfo's polymorphs.

We then use molecular-dynamics (MD) simulations with a linear heating
protocol to locate the phase transition temperature. We fix $k_{\rm
  tg}=2k_{\rm comp}=8$~eV/(\AA rad) for simplicity~\footnote{In a
simple landau model with a free-energy landscape of the form
$F(R)=-\frac{1}{2}\alpha R^2+\frac{1}{4}\beta R^4$, the curvature
around the minimum $R_0$ (that satisfies $\frac{\partial F}{\partial
  R}|{}_{R=R_0}=0$) is precisely $\alpha$, which is twice the
magnitude of the quadratic curvature of the double well at $R=0$.} and
vary the target tilt amplitude.  Starting from the $R3c$ phase of
\bfo\ in the orientation such that the polarization points in the
$[111]_{\rm pc}$ direction and the antiphase tilts are parallel to it,
we first thermalize the system for 20~ps at 1~K and, subsequently,
heat it at a rate of $10^{-4}$~K/fs until the transition to $Pnma$
occurs. Comparison with equilibrium simulations (see Supplemental)
shows that this protocol provides sufficiently converged transition
temperatures.

As a reference calculation, we first simulate bulk \bfo\ under the
epitaxial constraint imposed by a $(001)_{\rm pc}$-oriented SrTiO$_3$
substrate ($\sim$1\% compressive epitaxial strain). We use an
$8\times8\times8$ supercell, where the $1\times1\times1$ cell
corresponds to one perovskite formula unit. The $R3c$-to-$Pnma$
transition is identified at approximately 850~K. The in-phase tilt
develops in-plane, which reflects the symmetry breaking due to the
epitaxial strain.

\begin{figure}[t]
    \centering
    \includegraphics[width=1\linewidth]{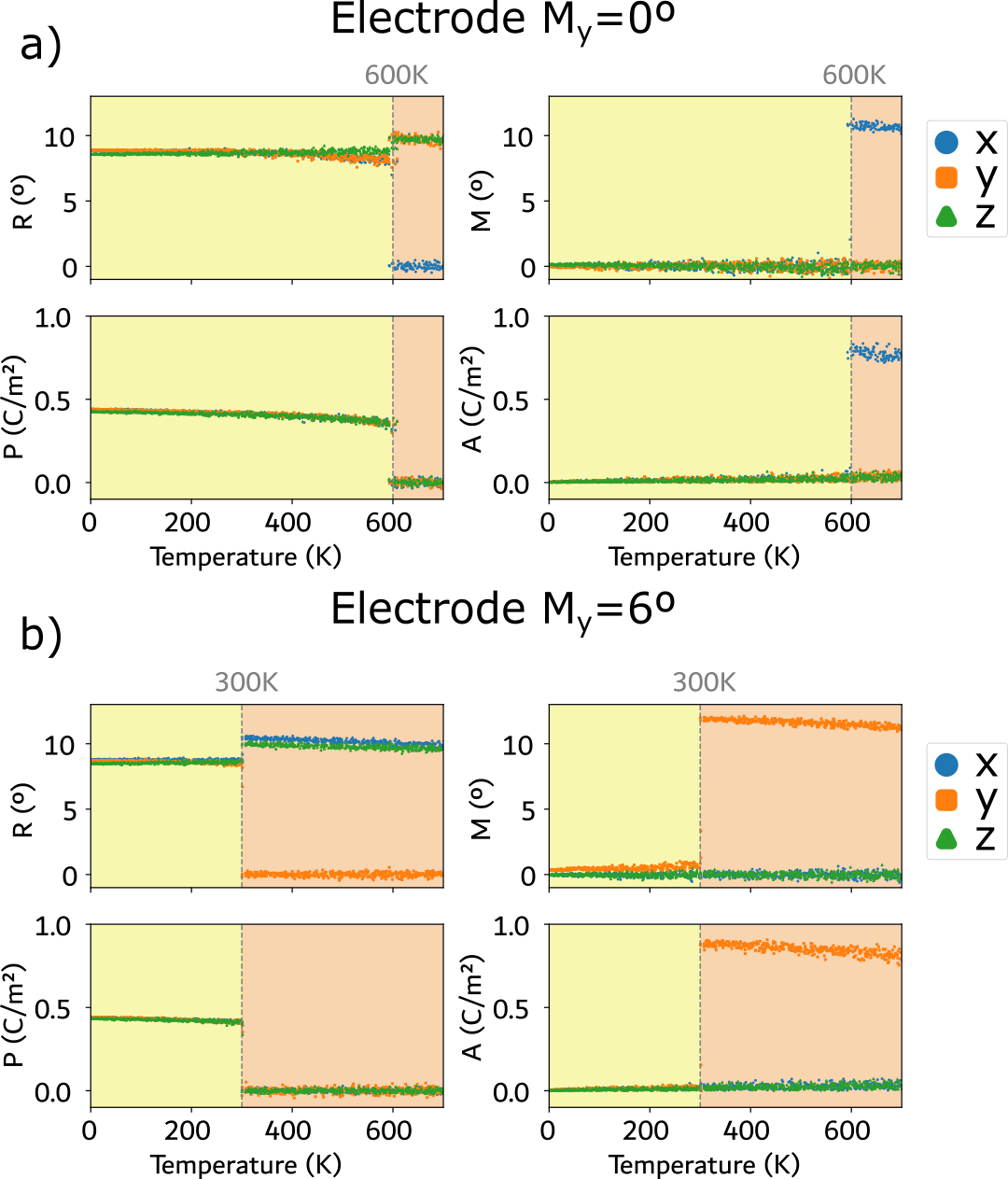}
    \caption{Temperature-driven phase transition of 3.2~nm
      \bfo\ nanocapacitors, with in-plane strain fixed to that of a
      $(001)_{\rm pc}$-oriented SrTiO${}_3$ substrate, for $(k_{\rm
        comp},k_{\rm tg})=(4,8)$~eV/(\AA rad). a) $f_Q=0$ (final
      $M_y=0^{\circ}$, cubic electrode) and b) $f_Q=1.2$ (final
      $M_y=6^{\circ}$, close to bulk SrRuO${}_3$'s tilts). The abrupt
      change from the $R3c$ polymorph (antiphase tilts and
      polarization fully developed) to the $Pnma$ polymorph (one
      antiphase tilt becomes in-phase, polarization is lost and
      antipolar mode develops) marks the transition temperature.}
    \label{fig:fig2}
\end{figure}

We next examine how the biased electrode modifies this transition in
symmetric \bfo\ nanocapacitors with ferroelectric thicknesses of
1.6~nm, 3.2~nm, and 6.4~nm. The simulated electrode is always 3.2~nm
thick, and the considered supercell comprises $8\times8$ perovskite
units in plane. We vary $f_Q$, producing electrode in-phase tilts
spanning $M_y\sim0^\circ$--$11^\circ$.

Figure~\ref{fig:fig2} shows the temperature evolution of the
structural order parameters for two contrasting cases: a cubic
electrode and an electrode whose realized tilt amplitudes are close to
those of bulk SrRuO$_3$. In both cases the transition to $Pnma$ is
marked by the emergence of in-phase tilts and antipolar distortions,
together with the loss of polarization. For the cubic electrode, the
two in-plane orientations for in-phase tilts are symmetry equivalent;
the tilt develops along $x$ in the simulation shown. By contrast, the
tilted electrode selects the orientation of the in-phase tilt in the
\bfo\ layer, which develops parallel to that in the electrode. Thus,
the electrode provides an orientational preference in addition to
modifying the relative stability of the competing phases.

\begin{figure}[t]
    \centering
    \includegraphics[width=1\linewidth]{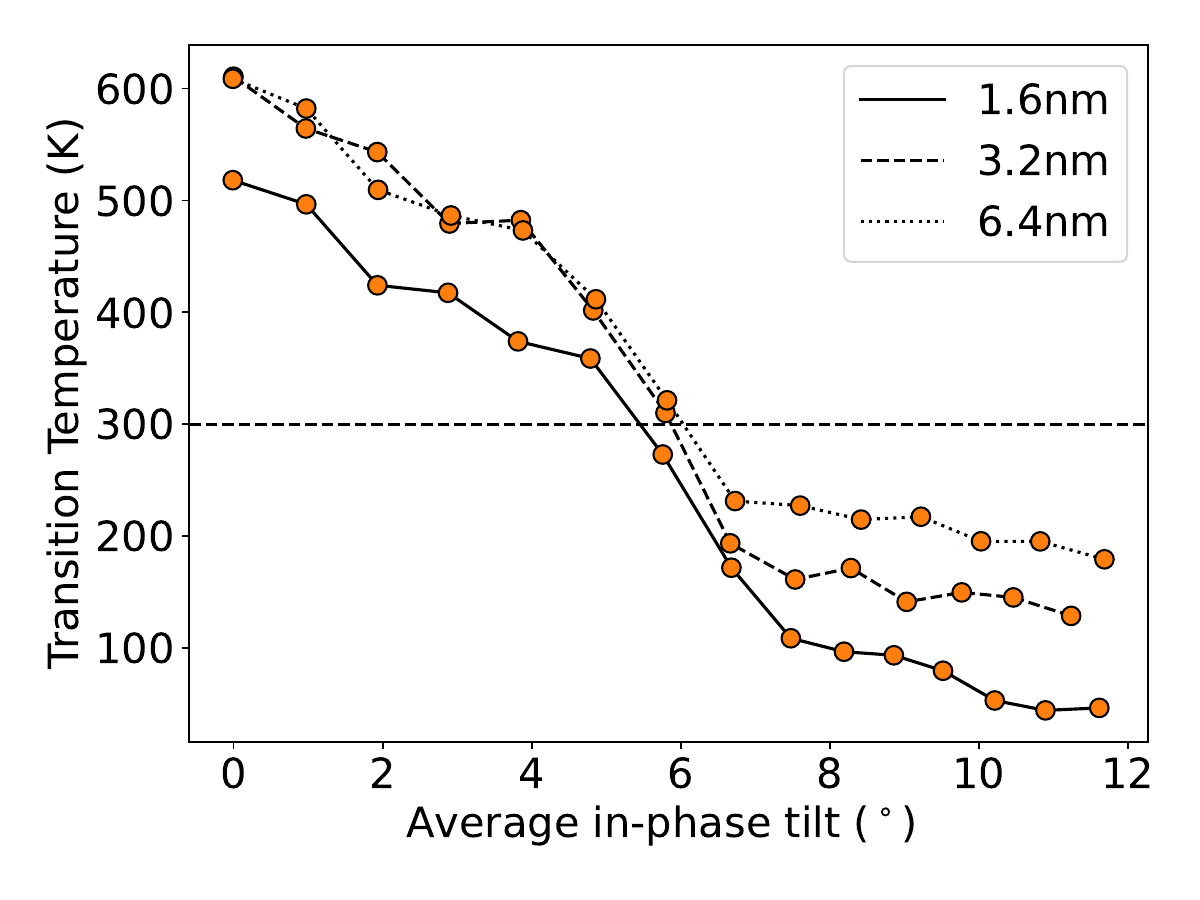}
    \caption{Computed $R3c$ to $Pnma$ transition temperatures for
      \bfo\ nanocapacitors of different thicknesses, as a function of
      electrode tilt amplitude ($a^-b^+a^-$ orientation).}
    \label{fig:fig3}
\end{figure}

Figure~\ref{fig:fig3} shows the computed $R3c$-to-$Pnma$ transition
temperature as a function of the realized electrode tilt amplitude for
the three ferroelectric thicknesses here considered. In the limit of
vanishing electrode tilt amplitude, the transition temperature is
close to the value obtained for strained bulk \bfo. (The difference
likely reflects the realistic treatment of screening in the
nanocapacitor simulations, which penalizes the $R3c$ phase compared to
the ideal short-circuit coniditions assumed in the calculations of
pure \bfo.) Increasing the electrode tilt amplitude systematically
lowers the transition temperature, consistent with an enhanced
preference for the electrode-favored $Pnma$ tilt pattern. This
reduction is more pronounced for thinner ferroelectric layers, which
have a larger interface-to-volume ratio, supporting the interfacial
origin of the effect. At sufficiently large tilt amplitudes, the
transition occurs below room temperature. In particular, for
amplitudes comparable to those of bulk SrRuO$_3$
($M_y\sim6^\circ$--$8^\circ$), all nanocapacitors considered here
stabilize in the $Pnma$ phase at room temperature.

The same interfacial bias that stabilizes the $Pnma$ phase at room
temperature may also reshape functional properties such as
electric-field-driven transitions.  We thus investigate how the
electrodes' phonon energy landscape affects field-driven switching in
the 3.2~nm \bfo\ nanocapacitor. This thickness is selected as a
representative intermediate case, for which the interfacial phonon
bias remains substantial while the ferroelectric layer is sufficiently
thick to be experimentally convenient. We focus on three final
electrode tilt amplitudes, $M_y=0^\circ$, $6^\circ$, and $9^\circ$.

\begin{figure}[t]
    \centering
    \includegraphics[width=1\linewidth]{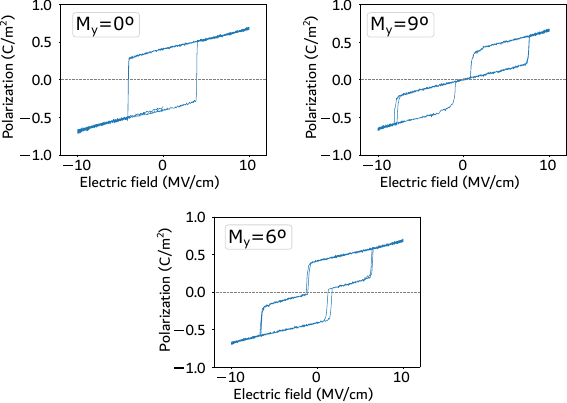}
    \caption{Calculated out-of-plane polarization--field loops for the 3.2~nm
      \bfo\ nanocapacitor with realized electrode tilt amplitudes
      $M_y=0^\circ$, $6^\circ$, and $9^\circ$ at 300 K.}
    \label{fig:fig4}
\end{figure}

To probe the resulting switching loops, we apply a sinusoidal electric
field along the vertical stacking direction, with a maximum amplitude
of 10~MV/cm and a frequency of 1~GHz, over two full cycles. The field is incorporated as effective forces exerted on
nominally charged ions in the ferroelectric layer ($+3$ for Bi and Fe,
$-2$ for O). The short time scales accessible to MD motivate the high
driving frequency and the use of large field amplitudes: field-driven
switching can involve thermally activated nucleation, and stronger
fields lower the relevant free-energy barriers, making these events
accessible within the simulated cycles. These calculations therefore
aim to understand qualitatively how the structure changes under an
applied electric field and how these changes depend on interfacial
phonon bias, rather than to predict experimental coercive fields
quantitatively. The resulting Polarization- Electric field ($P$--$E$) loops are shown in
Fig.~\ref{fig:fig4}.

In the limit of zero electrode tilts ($M_{y}=0^{\circ}$), the
nanocapacitor displays the usual ferroelectric loop. For the largest
$M_{y}$, the calculated $P$--$E$ curve displays the characteristic
double-loop shape of antiferroelectric switching. At zero field, the
nanocapacitor stabilizes in the antipolar $Pnma$ phase, whereas a
sufficiently large applied field induces the polar $R3c$ phase. The
reproducibility of this response over the simulated cycles indicates a
reversible field-driven transition between these structural phases.
For the intermediate electrode tilt amplitude, the two lobes of the
double loop become connected. In this case, the field-driven switching
pathway allows the nanocapacitor to transform from the polar $R3c$
phase to the antipolar $Pnma$ phase at comparatively small
fields. Once the nanocapacitor switches to $Pnma$, reducing the field
to zero, rather than continuing the cycle toward the next $R3c$
transition, allows the antipolar phase to be retained at zero
field. This behavior therefore suggests that, for electrodes with
intermediate tilt amplitudes, interfacial phonon bias can provide
access to a metastable antipolar zero-field state, potentially
enabling multistate functionality.

Interfacial phonon bias thus endows a structural dimension to
epitaxial stabilization beyond elastic and electric boundary
conditions. While epitaxial strain modifies phase stability by
constraining the avergage in-plane lattice parameters, coherent
interfaces impose further structural boundary conditions through the
connectivity of the atomic network. We expect this mechanism to be
most effective in heterostructures whose structural networks remain
connected across the interface, allowing distortions in one material
to influence those in the other. Perovskite oxides provide a natural
platform because their corner-sharing oxygen octahedra couple tilt
patterns across interfaces. Moreover, the strong (competitive)
coupling between polarization and octahedral tilts in most
perovskites~\cite{zhong95b,wojdel13,benedek13a} means that this
interfacial influence can extend directly to polar stability and
switching, making the mechanism relevant well beyond \bfo. For
controlling ferroelectric phase stability, particularly promising
systems are those in which polar and antipolar or nonpolar phases
compete and respond differently to the transmitted distortion. Beyond
perovskites, we expect the mechanism to apply whenever interfacial
bonding couples compatible distortions in adjacent materials.

Interface orientation provides an additional handle on this
interfacial coupling. For example, NdNiO$_{3}$ films grown on
NdGaO$_{3}$ exhibit a metal-to-insulator transition near 200~K for the
$(001)_{\rm pc}$ orientation, compared with 335~K for the $(111)_{\rm
  pc}$ orientation~\cite{Catalano_2018,10.1063/1.4919803}. This
orientation dependence has been attributed to substrate-imposed
octahedral tilts, illustrating how interfacial connectivity can
substantially modify a phase-transition temperature. (Each interfacial
octahedron shares one oxygen across a $(001)_{\rm pc}$ interface but
three across a $(111)_{\rm pc}$ interface, imposing stronger
constraints on the tilt pattern in the latter.) Further control could
be achieved by tuning the phonon energy landscape of the
substrate.

In practice, electrode composition, interface orientation, epitaxial
strain, layer thickness, and temperature offer distinct routes to tune
interfacial phonon bias. These parameters can modify the preferred
electrode distortion, its stiffness, and the extent of its influence
on the ferroelectric layer. In real heterostructures, such changes
will generally interplay with other boundary conditions. The
bias-potential approach used here provides a controlled model to
isolate the impact of the electrodes' internal energy landscape while
elastic, electric, and chemical constraints remain the same. This
separation can guide the choice of electrode materials and interface
geometries that favor a desired ferroelectric phase or switching
response. We hope that the striking behaviors revealed and explained
here (including the ability to induce distinct electrical responses in
\bfo\ nanocapacitors at ambient conditions) will stimulate further
theoretical and experimental exploration of this general design
strategy.

\section{Acknowledgements}
Work supported by the Luxembourg National Research Fund (FNR) through Grant No. BRIDGES/18421428/SWITCHON. U.D. also acknowledges FNR's support through Grant No. C23/MS/17909853/BUBBLACED. Work supported in part by Intel.

\bibliography{biblio}

\section{Machine-learned interatomic potential details}

\begin{figure}[t]
    \centering
    \includegraphics[width=1\linewidth]{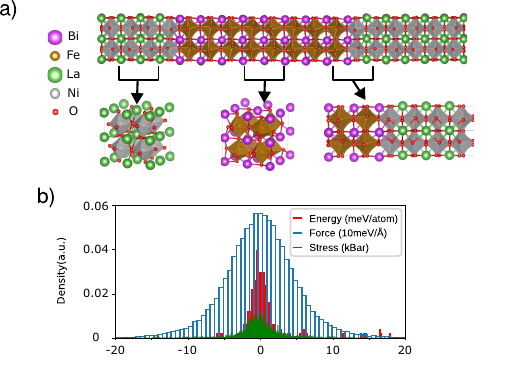}
    \caption{a) Illustration of nanocapacitor architecture and training structures emulating the different environments (crystal structure figures obtained using VESTA software \cite{vesta11}). b) Histogram of errors in energies, forces and stresses on the validation set.}
    \label{fig:supfig1}
\end{figure}

We construct the nanocapacitor machine-learned interatomic potential (MLIP) using the neuroevolution potential architecture as implemented in GPUMD \cite{GPUMD}. We gather the DFT data required for training (crystal structures with energies, forces and stresses) using VASP's on-the-fly algorithm.
The usual approach consists on running ab initio molecular dynamics (AIMD) of the structures to be trained on. In our case, the required size of the nanocapacitors makes it impossible to run AIMD of the full nanocapacitor. We go around this issue by ``breaking up" the nanocapacitor in parts. As the NEP model captures the energetics based on the environment of each atom, we can decompose these environments into three: bulk \bfo\, bulk \lno\ and interface (see Fig.~\ref{fig:supfig1}a). These cells are small enough that we can run AIMD on them and span the local environments of atoms in any nanocapacitor size. The bulk environment training was done using a $2\times2\times2$ perovskite supercell with AIMD at 300 K starting from the relaxed $R3c$ and $Pnma$ polymorphs of \bfo\ and R$\bar3$c polymorph of \lno, and it was extended with a high temperature training for bulk \bfo\ using a temperature ramp from 300 K to 1000 K. The training on the interface environment was done at 300 K, starting from the relaxed structure of two different orientations of \bfo\, with the polarization along the $[111]_{\rm pc}$ (referref to as the $1^+$ configuration) and along $[11\bar1]_{\rm pc}$, with the anti-phase tilts (R) parallel to the polarization. We extended the training of the nanocapacitor by training in the presence of an effective constant negative electric field, starting from the $1^+$ configuration, with amplitudes 2, 4 and 6 MV/cm. In this way, we can explore the potential well away from the minimum towards switching. We gather a total of 2426 structures which we use to train the neuroevolution potential (NEP) MLIP (training and validation data are available in Zenodo \footnote{All data ara available at 10.5281/zenodo.22769801}), split into 1436 of bulk \bfo\, 75 of bulk \lno\ and 915 of the nanocapacitor.

To validate the model, we extract 100 equispaced snapshots from all the different AIMD simulations and run single-point DFT calculations. These calculations are not used to train the model, thus they serve as a measure of the goodness of the model to interpolate. We show the histogram for the errors in the prediction of energies, forces and stresses in Fig.~\ref{fig:supfig1}b. The root mean squared errors (RMSE) for the validation set are 5.4 meV/atom for the energies, 64 meV/\AA\ for the forces and 4.2 kbar for the stresses, well within the usual errors reported in the literature.

We are mainly interested in the structural properties as a function of temperature. We thus further test the model using bulk \bfo\ for benchmarking purposes as described in the main text.
Notice that the transition to the $Pnma$-like polymorph happens with the in-phase tilt axis in-plane. This implies that this configuration has lower energy than the one with in-phase tilt out-of-plane. We confirm that this is the case by comparing the relative energy of both configurations after a full relaxation. The MLIP predicts this energy difference to be 16 meV/fu, while DFT predicts 18 meV/fu. Not having trained on any constrained configuration, this further validates that our model performs well at extrapolating. From these tests we conclude that the MLIP for the nanocapacitors is accurate both for interpolation and extrapolation.

In this work we opted for symmetric nanocapacitors. First, by symmetric we refer to nanocapacitors where both interfaces are chemically equivalent. A usual way of understanding the perovskite structure with ABO$_3$ chemical formula is as a stacking of alternating AO and BO$_2$ layers. Constructing a superstructure mixing two different perovskites, ABO$_3$ and A'B'O$_3$, results in two interfaces, BO$_2$/A'O and B'O$_2$/AO, which are usually different. This is the source of interesting phenomena like that of 2 dimensional electron gases (2DEGs) at the interface of differently nominally charged layers as in, e.g., LaAlO$_3$/SrTiO$_3$ superstructures. Even in systems where there is no charge discontinuity in neighboring layers, this asymmetry always introduces a built-in field which makes nanocapacitor states with polarization up and down (in the direction orthogonal to the interface) different, making it harder to extract what may be subtle energy differences due to other effects, such as the ones we are focusing on in this work. A symmetric interface would be one where we add one more layer of one of the perovskites, breaking stoichiometry, but removing the built-in fields from interface asymmetry. In what follows, we symmetrize the interface by adding one extra BiO layer, so that our symmetric interfaces are BiO/NiO$_2$.

\section{Definition of structural order parameters}
The structural order parameters are extracted from each MD snapshot by projecting atomic displacements onto a set of symmetry-adapted distortion modes defined for the $2\times2\times2$ perovskite reference unit cell. For each snapshot, the fractional displacement of every atom $i$ from its reference position is computed as $\boldsymbol{\delta}_i = \mathbf{u}_i - \mathbf{u}_i^{(0)}$. We use the usual cubic perovskite positions as reference. The six octahedral tilt amplitudes --- three R-point ($R_x$, $R_y$, $R_z$) and three M-point ($M_x$, $M_y$, $M_z$) ---, and the 6 antipolar modes are obtained as inner products of the displacement field with the corresponding symmetry-mode vectors. Since antipolar modes modulated along $X$, $Y$ and $Z$-points are degenerate, we plot the amplitude of each pair of modes. This amplitude is then transformed to physical units (C/m${}^2$) using the average displacement per Bi atom and nominal charges. The polarization ``proxy" is computed as
\begin{equation}
    \mathbf{P} = \frac{f_{\mathrm{bulk}}}{N_{\mathrm{f.u.}}} \sum_i Z_i\, \boldsymbol{\delta}_i,
\end{equation}
where $Z_i$ are isotropic nominal charges ($Z = +3$ for Bi, Fe, $Z = -2$ for O), $N_{\mathrm{f.u.}}$ is the number of formula units, and $f_{\mathrm{bulk}}$ is a conversion factor to physical units (C/m$^2$). Even though this is a rough approximation, it is known to work well for \bfo. 


\section{Computational methods}\label{sec:supcm}
\begin{figure}
    \centering
    \includegraphics[width=1\linewidth]{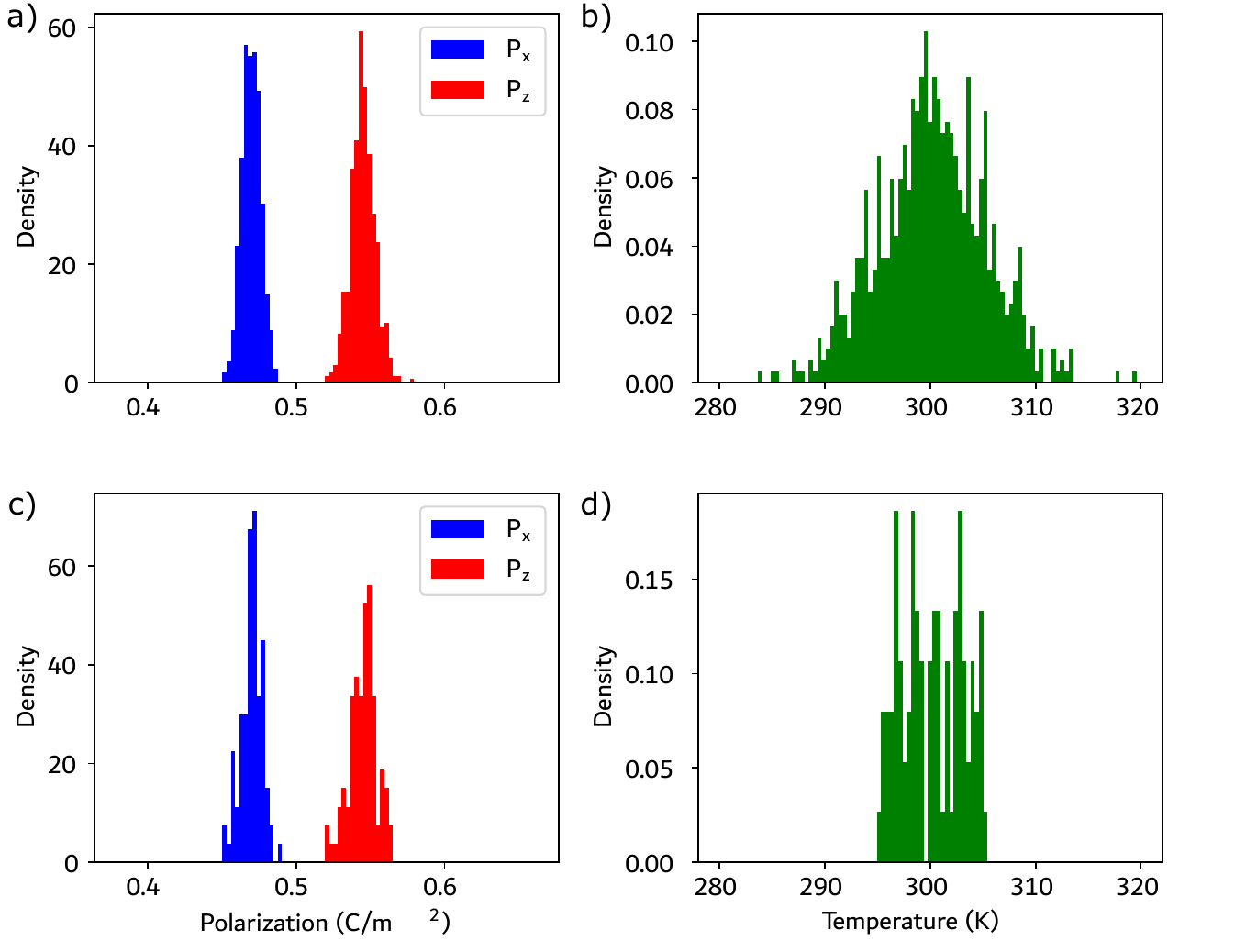}
    \caption{Comparison between usual equilibrium MD methods and the temperature ramp approach. a) Polarization histograms for the constant temperature run. b) Temperature histogram for the constant temperature run. c) Polarization histogram for a similar temperature window in the temperature ramp protocol. Temperature histogram of the same temperature window as in c). d) Temperature histogram for the temperature ramp approach.}
    \label{fig:supfig2}
\end{figure}

All first-principles calculations were performed using the projector augmented-wave (PAW) method \cite{blochl94} as implemented in VASP \cite{kresse96,kresse99}, with the PBEsol exchange-correlation
functional \cite{perdew96}. We solve explicitly for 6 electrons of O (2s${}^2$p${}^4$), 15 of Bi (5d${}^{10}$6s${}^2$6p${}^3$), 14 of Fe (3p${}^6$3d${}^7$4s${}^1$), 16 of Ni (3p${}^6$3d${}^9$4s${}^1$) and 11 of La (5s${}^{2}$5p${}^6$5d${}^1$6s${}^2$). The plane-wave energy cutoff was set to 500~eV in all cases. We use a k-point grid of $2\times2\times2$ for bulk \bfo, $4\times4\times4$ for bulk \lno, and $4\times4\times1$ for the nanocapacitors. We add a Hubbard U correction of U=4 to the Fe atoms as usual for \bfo\ magnetic calculations, and the magnetic state of \bfo\ is always converged to the G-type antiferromagnetic state.

The molecular dynamics simulations were performed using GPUMD software. We use the Martyna-Tuckerman-Tobias-Klein ensemble \cite{MTTK}, with a Nosé-Hoover thermostat \cite{Hoover} and a Parrinello-Rahman barostat \cite{ParrinelloRahman}. We modify the GPUMD code to fix the in-plane strain and allow for the third lattice vector to move according to the barostat. We use a timestep of 1 fs and a constant temperature ramp of 0-1000 K during 10 ns, which represents a heating of $10^{-4}$ K/fs. We set the thermostat coupling to default and the barostat period to 10000 fs.

The phase transition we are targeting is between \bfo\ $R3c$ and $Pnma$ polymorphs. The $R3c$ polymorph has a polarization in the $[111]_{\rm pc}$ direction, with anti-phase tilts parallel. The $Pnma$ phase, has anti-phase tilts in the $[101]_{\rm pc}$ direction, in-phase in the $[010]_{\rm pc}$ and an antipolar distortion modulated parallel to the in-phase tilts, with no net polarization. Since it is a first order phase transition, we can easily capture the transition temperature by computing a running average of the polarization and finding the abrupt change from finite average polarization to 0. We checked several values for the running average window and threshold for the transition temperature determination, with minimal impact on the numerical value of the transition temperature itself.

One may argue that our MD simulations are never at equilibrium, due to the slow heating up protocol we follow to capture the transition temperature. On the one hand, this method should only overestimate the transition temperature due to not spending enough time at a particular temperature. In that sense, our results serve as a higher bound for the transition temperature. We run a further test comparing two MD simulations, a usual fixed-target-temperature at 300 K run for 10 ns (see Fig.~\ref{fig:supfig2}a and b) and our temperature ramp protocol (see Fig.~\ref{fig:supfig2}c and d). From the constant temperature run, we extract the average ($T_{avg}$) and standard deviation ($\Delta T$) of temperatures. We use them to extract a window of data from the ramp protocol from $T_{avg}-\Delta T$ to $T_{avg}+\Delta T$. We compare the resulting histograms in Fig.~\ref{fig:supfig2}. We can see that the temperature histogram for the temperature ramp protocol is basically flat due to the constant heating and the polarization histograms are very similar, with an error of less than 0.03\% in the average value.


\section{Determination of $k_{\rm comp}$}

\begin{figure}
    \centering
    \includegraphics[width=1\linewidth]{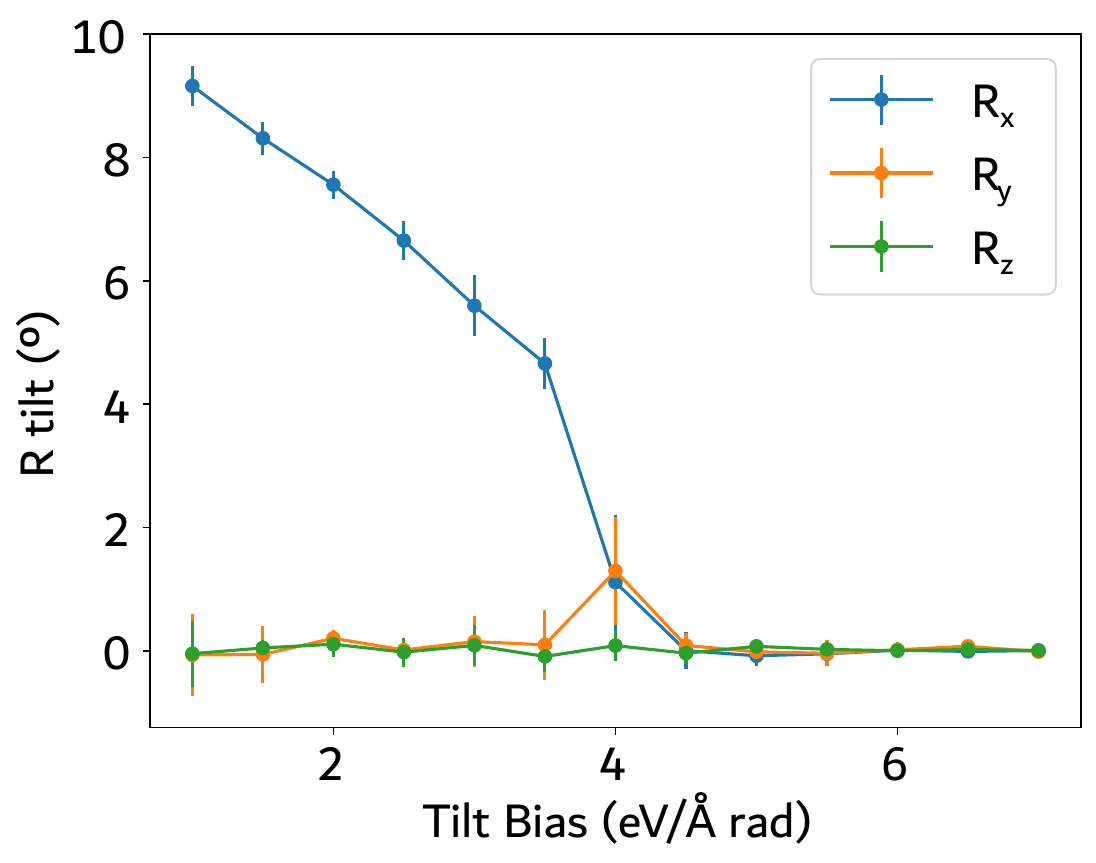}
    \caption{R tilt average amplitude and standard deviation for different tilt bias amplitudes. We can see how the transition between tetragonal and cubic lies around k= 4eV/(\AA rad).}
    \label{fig:supfig_n}
\end{figure}

We determine the amplitude of harmonic bias required to ``flatten" the energy landscape of \lno\ by running MD simulations at 10 K on a $2\times2\times2$ supercell with a time step of 1fs, and we increase the tilt bias by 0.5 eV/(\AA rad) every 10 ps (see Fig.~\ref{fig:supfig_n}). We can see clearly the transition between the native tetragonal structure ($k_{\text{comp}}<4$) and the target cubic structure ($k_{\text{comp}}>4$). We thus choose $k_{\text{comp}}=4$ eV/(\AA rad).




\end{document}